# Modeling Sheep pox Disease from the 1994-1998 Epidemic in Evros Prefecture, Greece

C. Malesios[1], N. Demiris[2], Z. Abas[3], K. Dadousis[4] and T. Koutroumanidis[5]

1. (Corresponding Author) Department of Agricultural Development, Democritus University of Thrace, Pantazidou 193, Orestiada, Tel/Fax: +30 25520 41133,
e-mail: malesios@agro.duth.gr
2. Department of Statistics, Athens University of Economics and Business, 76, Patission Str., Athens
3. Department of Agricultural Development, Democritus University of Thrace, Pantazidou 193, Orestiada
4. Rural Economy and Veterinary Directorate, Region of Eastern Macedonia and Thrace, Regional Unit of Evros, Orestiada
5. Department of Agricultural Development, Democritus University of Thrace, Pantazidou 193, Orestiada

## Abstract

Sheep pox is a highly transmissible disease which can cause serious loss of livestock and can therefore have major economic impact. We present data from sheep pox epidemics which occurred between 1994 and 1998. The data include weekly records of infected farms as well as a number of covariates. We implement Bayesian stochastic regression models which, in addition to various explanatory variables like seasonal and environmental/meteorological factors, also contain serial correlation structure based on variants of the Ornstein–Uhlenbeck process. We take a predictive view in model selection by utilizing deviance-based measures. The results indicate that seasonality and the number of infected farms are important predictors for sheep pox incidence.

# 1. Introduction

Sheep pox is a highly contagious viral infection of sheep (*Garner et al., 2000*). The causative agent belongs to the genus Capripoxvirus, one of the six genera of poxviruses of vertebrates (*Yeruham et al., 2007*). The causative virus of sheep pox is antigenically and genetically closely related to goat pox virus and lumpy skin disease virus. The disease spreads by the direct contact with infectious animals and indirect contact with contaminated objects; for example, the virus could survive for many years in dried scabs at ambient temperatures and for 2 months on wool (*OIE, 2002*). Infected sheep may transmit the virus (or the causative agent) at every stage of the disease, even up to eight weeks after the lesions have resolved (*Singh et al., 1979*). An additional transmission route is through biting flies (*Kitching and Mellor, 1986*). The transmission of sheep pox between flocks occurs when sheep are moved from flock to flock. This type of movement occurs frequently in the Middle East (*DEFRA, 2012*) (For a comprehensive review on the history and distribution of the disease worldwide see *Rao and Bandyopadhyay (2000)* and *Bhanuprakash et al. (2006)*).

The disease may spread rapidly, resulting in the fatality of infected animals within a few days. This form of the disease is seen mostly in lambs. Characteristic symptoms are fever and paralysis. An eruption in the form of red spots appears on the membranes of the eyes and nose, and on the wool-free parts of the skin. In older sheep, the disease begins by signs of serious ill-health, notably a high temperature and suppressed appetite (*DEFRA, 2012*). Infected pregnant ewes often abort. The disease is characterized by high mortality rates and significant decreases in milk production (*Yeruham et al., 2007; Belwal et al., 1982*). Economic losses due to mortality and milk production losses are also noticeable (*Senthilkumar et al., 2010; Garner et al., 2000*).

Another characteristic of the disease noted in the relevant literature is the one of seasonal variations of outbreaks of sheep pox. Specifically, most studies associate sheep pox occurrence with unusual winter conditions and report that the majority of outbreaks occur during the winter and spring months (e.g. *Hailat et al., 1994; Yeruham et al., 2007; Bhanuprakash et al., 2005*). There are also various environmental/meteorological factors which can influence disease occurrence, such as the amount of rainfall, relative humidity and maximum temperature (e.g. *Bhanuprakash et al., 2004; Bhanuprakash et al., 2005; Webster, 1981*).

Sheep pox is widely distributed in Asia and North and East Africa. Most of Europe and the Americas are now free from endemic sheep pox. However, the disease has occurred in Greece in 2000 (only 1 infected herd). Greece has a history on sheep pox appearances in the recent years. Specifically, in 1987 there was an incidence of sheep pox in the island of Lesbos which consisted of four outbreaks. The disease was controlled via culling of infected animals and vaccination of the neiboring farms within the protected zone (*Mangana et al., 2008*). Re-appearance of the disease occurred with a single outbreak in the Evros Prefecture during 1988. This last case of sheep pox in the Evros Prefecture, before the major outbreak of 1994, was controlled through the slaughtering of infected flocks and vaccination of all sheep farms near the Evros river. 1992 was a turning point regarding the implementation of control policies aimed at eradicating sheep pox outbreaks in Greece. In particular, a stamping out/non-vaccination policy replaced policies which up to then included, in addition to slaughter of infected animals, vaccination of the neighboring farms for controlling the disease. The new control measures also included the cleaning and disinfection of culled premises and the establishment of protection and surveillance zones of a radius of 3 and 10 km, respectively, around the outbreak.

A major outbreak appeared in Evros in 1994, and over the next four years several outbreaks occurred in Evros and Thessaloniki (1995), Larissa, Xanthi, Rhodopi, Kavala, Magnissia, Evros and Lesbos in 1996, Kavala, Magnissia, Halkidiki, Evros and Rhodopi in 1997. A decline of sheep pox outbreaks was eventually achieved in 1998, with outbreaks restricted only in the Evros Prefecture. Following an absence of appearance in 1999, sheep pox has re-appeared in 2000 (only a single incidence).

During 1996 we observed the highest frequency of sheep pox incidents in Greece. This high rate of appearance of sheep pox is attributed to various factors, such as efficient reporting of outbreaks at that time, a higher number of susceptible animals and host/agent factors *(Mangana et al. 2008).*

Evros Prefecture is a critical area as regards the appearance of contagious animal diseases on European soil (another example is foot-and-mouth disease (FMD)). This is due to that it constitutes the natural passage for the transfer from Asia to Europe of infectious diseases endemic in their regions of origin (*EFSA, 2006*). Sheep pox caused serious economical losses to the livestock of Evros Prefecture. During the last major sheep pox epidemic in 1998, eradication measures have been implemented to control the disease. Stamping out of the animals from affected flocks and flock vaccination for the risk areas have been implemented. As *Mangana et al. (2008)* stress, the geography of the region, with river Evros constituting the natural border between Greece and Turkey and Bulgaria, makes it easy for sheep in the Greek soil to come in close contact with scabs from dead animals that could transmit the disease. Another possible reason could be attributed to the movement of workers and farmers from neighboring countries (i.e. Turkey and Bulgaria), where sheep pox outbreaks are frequent. *Dadousis (2003)* also relates sheep pox outbreak in Greece with Turkey based on the evidence of high incidence rate in Turkey and the 1996-1997 epidemic peak in Evros, Greece. There is additional evidence to believe that pox virus was introduced in Greece from Turkey, since the disease is endemic in Turkey and the primary incidents have been detected close to Evros river which mostly determines the border with Turkey (*Dadousis, 2003*).

The aim of the current study was to examine the dynamics of the spread of sheep pox and to address the various epizootiology issues described previously (e.g. temporal and climatic factors) in this sensitive region. In doing so, we implement Bayesian stochastic regression models to describe sheep pox occurrences in Evros during 1994-1998. A stochastic regression model differs from a standard regression model in that the former allows for one or more of the explanatory variables, such as $U_i$, to be random. A number of explanatory variables were utilised, including seasonal and environmental/meteorological factors, in order to assess the magnitude of their effect on the occurrence of the outbreak. In addition, serial correlation structure, based on variants of the Ornstein–Uhlenbeck process represents an integral part of our modeling approach.

**2. Materials and Methods**

*2.1 Data*

The sheep pox epidemics of the Evros Prefecture of Northeastern Greece began on November 1994, ended in December 1998 and included 249 infected premises (a re-appearance of the disease with only one infected herd occurred in 2000). The course of the epidemic from 1994 up to 2000 is depicted in the following map (Fig. 1).

**INSERT FIGURE 1 APPROXIMATELY HERE**

Farm-level data for the sheep pox epidemic were obtained from the Veterinary Directorate of Northern Evros Prefecture *(VDNEP)*, located in the city of Orestiada, Evros Prefecture, Greece. Treating each infected farm as a single unit, the data comprised of daily records on the number of infected farms. Temporal information, important for identifying the progress of disease through time was also made available, such as the exact day of the putative infection time (as determined by the examination of the infected animals in each farm by the local veterinary services). The daily records were transformed to weekly records to avoid substantial uncertainties related to the exact day of infection. Spatial information concerning the location of each farm was not available. However, we obtained a crude approximation of the farm location by the village wherein each farm is located. Following *Choi et al. (2012)*, who model foot-and-mouth disease in cattle, we also include an epidemic-type component, the number of villages with sheep pox in the previous week.

Finally, the acquirement of 1994-98 meteorological data from the Greek National Meteorological Service (http://www.hnms.gr/hnms/english/index_html) for the Evros Prefecture region, where the epidemic was located, enables the assessment of their significance through statistical modeling. The following meteorological variables were included as predictors: (a) rainfall, (b) average temperature, (c) max temperature, (d) min temperature, (e) humidity.

This dataset has not been previously presented or analyzed and presents an opportunity to obtain insight regarding the patterns of spread in such diseases.

*2.2 Modeling*

We implement Bayesian regression models inspired by *Choi et al. (2012)* and compare the various models derived through deviance-based measures. We have chosen to model weekly data on the sheep pox cases as opposed to monthly counts for finer resolution and accuracy. For the purposes of our analysis, a case is a farm infected with sheep pox virus. We fitted five Bayesian regression processes for modeling the response counts, corresponding to suitable distributions for count data, such as the Poisson and the negative binomial (NB). The latter is often used as an overdispersed alternative of the Poisson model. We do not utilize continuous sampling distributions – as in *Choi et al. (2012)* where a Gaussian and a square-root transform model were used – due to the problems arising when using continuous sampling distributions for datasets consisting of point observations (*Fernandez and Steel, 1998*). A slightly simpler version of this model was used in *Branscum et al. (2008)*.

In real life settings like the epidemic studied here, an inherent feature of the data is the excessive number of zero responses - data of this type are often referred to as zero inflated. Zero-inflated models for count data (see *Lambert, 1992*) have been widely used in statistics (see, e.g., *Böhning et al., 1999; Hall, 2000*). The zero-inflated distribution D of a random variable Y~ZID($p,\theta$) has a probability density function of the form:

$$f_{ZID} = pI_{\{y=0\}} + (1-p)f_D(y;\theta),$$

where $p$ is an additional parameter denoting the proportion of excess zeros in the data y, $f_D$ denotes the probability of distribution D and $I$ the indicator function. It is easy to see that the probability of a zero count is $p+(1-p)f_D(0|\theta)$. We have implemented the zero-inflated Poisson (ZIP) and zero-inflated negative binomial (ZINB) distributions linking covariates through the Poisson and negative binomial rates as well as the zero excess probability.

*2.2.1 Model Class*

Let $y_i$ denote the number of farms infected with sheep pox virus at time $t_i$, where $i \in \{0,1,...,259\}$ correspond to the time after $s$ years and $w$ weeks since the beginning of the study, where $s \in \{0,1,2,3,4\}$ and $w \in \{0,1,2,...,51\}$. The standard Poisson and negative binomial models are summarized as:

$$Y_i \sim f(y_i | \theta_i)$$

$$\theta_i = h(\mu_i)$$

$$\mu_i = \mathbf{X}_i^t \cdot \boldsymbol{\beta} + b_i + U_i + \tau \cdot y_{i-1},$$

where for the Poisson with parameter $\lambda_i$ it holds that $\mu_i = \log(\lambda_i)$ while for the negative binomial with parameters $r$, $q_i$ we have $E(y_i) = \mu_i = \dfrac{r(1-q_i)}{q_i}$.

Additionally, we may express the Bayesian zero-inflated Poisson (ZIP) and zero-inflated negative binomial (ZINB) regression models as a two-component mixture model as follows:

$$Y_i \sim \Phi(y_i | \theta_i; p_i)$$

$$\Phi(y_i \mid \theta_i; p_i) = p_i I_{\{y_i=0\}} + (1-p_i) f(y_i \mid \theta_i)$$

$$\theta_i = h(\mu_i)$$

$$\mu_i = \mathbf{X}_i^t \cdot \boldsymbol{\beta} + b_i + U_i + \tau \cdot y_{i-1},$$

where $I_{\{y_i=0\}}$ is an indicator variable for whether or not the response is positive, and the probability of excess zeros $p_i$ is a mixture proportion with $0 \leq p_i \leq 1$.

Finally, in addition to regressing the Poisson/negative binomial component of the models to covariates, one could try to similarly link the probability of excess zeros to covariate information, adding the following additional equation:

$$\log\left(\frac{p_i}{1-p_i}\right) = \mathbf{X}_i^t \cdot \boldsymbol{\beta}^z + b_i^z + U_i^z + \tau^z \cdot y_{i-1}.$$

The two sets of covariates may or may not coincide. In most occasions the set of covariates for both Poisson rate and zero excess probability is chosen to be a priori the same.

In the previously described models, $U_i$ is the stochastic Ornstein–Uhlenbeck process for modeling serial correlation between the counts of sheep pox cases, with mean zero and correlation given by $Corr(U_i, U_j) = \rho^{|t_i, t_j|}$, where $\rho$ denotes the correlation between two consecutive weeks (see *Choi et al., 2012*), and this can be thought of as a continuous-time analog of standard AR(1) models. The $b_i$'s are independent random effects associated with the $i^{th}$ year, and $\tau$ represents the parameter associated with the influence of the number of sheep pox cases in the previous week. We denote by $\boldsymbol{\beta} = (\beta_0, \beta_1, \beta_2, \beta_3, \ldots, \beta_9)^t$ the vector of regression coefficients for the intercept ($\beta_0$), and covariates describing number of villages infected in the previous week ($\beta_1$), average rainfall ($\beta_2$), average temperature ($\beta_3$), average maximum temperature ($\beta_4$), average minimum temperature ($\beta_5$), average humidity ($\beta_6$) and seasonal effects [spring ($\beta_7$), summer ($\beta_8$), autumn ($\beta_9$)].

*2.2.2 Prior Specification*

As an alternative to the independence assumptions considered between the meteorological fixed-effects covariates in other relevant studies (e.g. *Choi et al, 2012*), we adopt a *g*-prior type of approach for the specification of the prior densities of parameters of interest (i.e. meteorological and other fixed-effects covariates) to account for potential correlation among them. Specifically, we utilize Zellner's informative *g*-prior (*Zellner, 1986*) appropriately adjusted for generalized linear models (*Bové and Held, 2011*), which although sets a priori the covariates to be centered at zero, it takes the dependence of the covariates into account allowing for the weighting of the prior information contributing to the posterior of each parameter depending on the choice of *g*.

Suitably vague priors for the parameters of the Poisson, negative binomial, ZIP and ZINB models have been specified. The prior distributions for the fixed-effects parameters (**β**) under the *g*-prior specification approach involved the multivariate Gaussian distribution, with zero-mean vector and a prior variance of the form: $\frac{g}{\phi}(\mathbf{X}^t\mathbf{X})^{-1}$. For the selection of *g* we follow the *Kass and Wasserman (1995)* unit information prior which assigns *g*=n where n denotes sample size (here n=260). We have also conducted a sensitivity analysis over different values for *g* and obtained similar results. The choice of the precision parameter $\phi$, depends on the GLM model specification and we adopt a minimally empirical Bayes approach by setting the dispersion parameter to be the median of its estimates after having seen the data. Specifying this prior is necessary when moving beyond the Gaussian assumption, see

*Bové and Held (2011)* for the details. A weakly-informative prior $N(0,10^4)$ was specified for the $\tau$ parameter associated with the number of sheep pox cases in the previous week. Random year effects ($b_i$) are assumed to be normally distributed with zero mean and their standard deviation following a uniform $U(0,100)$ distribution (*Gelman, 2006*).

*2.2.3 Inference and Model Selection*

For point and interval parameter estimation we adopted the Bayesian paradigm, and we have used WinBUGS (*Lunn et al., 2000*) to fit the models via Markov chain Monte Carlo techniques (*Gelman et al., 2003*). The posterior summaries have been obtained by using 5,000 iterations as the burn-in period and an additional sample of 10,000 iterations (with thinning one out of ten iterations in order to reduce the autocorrelation of the samples for some parameters).

In this paper we take a predictive view to model selection. Therefore, we resort to deviance-based measures due to the well-known equivalence in model selection using cross-validation or AIC (see *Stone, 1977*). Thus, since there is no formal equivalence proof for models with complexity similar to those entertained in the present paper (cf. the discussion of *Spiegelhalter et al., 2002*) we chose to perform model selection based upon the mean deviance as well as the deviance information criterion (DIC) (*Spiegelhalter et al., 2002*), with smaller values being preferable. The WinBUGS code of the fitted models and the sheep pox data are available upon request by the corresponding author.

## 3. Results

*3.1. Descriptive analysis on the 1994-1998 Sheep pox Epidemic in Evros Prefecture, Greece*

The case index during this epizootic outbreak was on November 2, 1994, in a farm of Fylakio village, at a distance of 12 km from the Evros river on the border with Turkey. This represents the only case of disease occurrence at some distance from the Turkish border. The 83 animals of the farm were culled, as well as the sheep of small neighboring farms (*Official records of the VDNEP*). On October 17, 1995 there was re-appearance of sheep pox in the village of Petrades, located very close to the Evros river. The following cases were in the village of Thourio, in the plain of Orestiada, in the village of Vyssa and in the village of Asimenio. All the villages are near the river Evros, the natural border with Turkey.

Figure 2 displays the temporal distribution of the sheep pox cases - treating the farms as single units.

**INSERT FIGURE 2 APPROXIMATELY HERE**

Table 1 shows the total number of sheep pox cases and total number of dead sheep during the 1994-98 period in the Evros Prefecture, Greece.

**INSERT TABLE 1 APPROXIMATELY HERE**

Summarizing the above tables, 83 and 677 animals were culled during 1994 and 1995 respectively, whereas 1996 was the year with the highest impact and 22,710 dead animals. Subsequently, the numbers declined somewhat, with 8,948 and 3,024 dead animals in 1997 and 1998, respectively. Accordingly, cases of sheep pox were higher in 1996 (135), followed by 1997 (54 cases) and 1998 (48 cases). There were only two cases during 1994, and eight during 1995. After the sheep pox-free year of 1999, there was only one incidence of sheep pox at a farm in the planes of the city of Orestiada, which resulted in the culling of the 507 animals of the farm. In this analysis we exclude this event and restrict ourselves to the consecutive years between 1994 and 1998 to examine the progress of the disease.

An initial inspection of the data indicates that the disease peaked during the autumn season, with 62 cases in September, 64 cases in October and 66 cases in November. The remaining cases appeared during the summer months except June (July: 18 cases, August: 18 cases), and winter except February (January: 1 case, December: 20 cases). No cases were reported during spring. Therefore, the highest death tolls are observed during autumn, particularly November, with a total of 9,862 slaughtered animals, followed by October (9,540 dead animals) and September (6,483 dead animals). December is the year with the next higher death rate (4,801 animals), followed by August (1,362 animals) and July (1,284 animals). Finally 110 animals were culled during January. With regard to the seasonality of incidents, it is observed from empirical studies that from summer until January there was an increase of outbreaks of sheep pox.

*3.2. Epidemic analysis*

Posterior estimates (i.e. posterior medians along with the corresponding 95% credible intervals) of the parameters of interest from the Bayesian models were obtained, including estimates of variance component parameters. Specifically, we have fitted the Poisson, NB, ZIP and ZINB models, where for the zero-inflated models the probability of excess zeros is not regressed upon covariates (denoted by ZIP and ZINB). We additionally fit a ZIP model (denoted by $ZIP_h$) where, in addition to the Poisson component, we link excess zero counts to covariates.

Model selection criteria for the five models are presented in Table 2. We present the posterior mean deviance, the effective number of parameters and the deviance information criterion (DIC).

**INSERT TABLE 2 APPROXIMATELY HERE**

Regarding model fit, we observe that the Poisson and zero-inflated Poisson models provided the best fit to the data, as indicated by the values of the fit statistics (deviance = 310.2, 277.3 and 270, respectively).

Between the Poisson and NB models, the former performs better (lower DIC), in spite of the penalty imposed due to the larger $p_D$ value, for the number of parameters ($p_D$ = 56.52 and 45.36 for Poisson and NB, respectively). However, as shown by figures 3(a) and 3(b), the Poisson model produced partly inadequate estimates, in the sense that the agreement between the observed and the predicted sheep pox cases was imperfect. In contrast, the $ZIP_h$ model that additionally accounts for excess zero counts, had the best fit to the data.

Figures 3(a) to 3(c) present observed counts vs predicted counts for the three best models (i.e. Poisson and the two ZIP models). The models appear to achieve a satisfactory fit to the observed sheep pox count occurrences with a slight tendency to underestimate in some instances the phase of the epidemic. The worst prediction, especially concerning the last major peak of the disease is observed in the case of the ZIP model (figure 3b) where uncertainty is large. The $ZIP_h$ model seems to predict better the progress of the disease achieving a better prediction of all peaks of the disease (Figure 3c).

**INSERT FIGURE 3(a) APPROXIMATELY HERE**

**INSERT FIGURE 3(b) APPROXIMATELY HERE**

**INSERT FIGURE 3(c) APPROXIMATELY HERE**

Hence, conditional upon the selection of the best distributional form, based on deviance, we do covariate selection based on eliminating the least significant covariate to conclude with the best model for explaining sheep pox disease occurrence. Table 3 presents the results of the final selected $ZIP_h$ model (hence we present the posterior medians for the parameters for the Poisson rate and the zero excess probability along with the 95% posterior intervals).

**INSERT TABLE 3 APPROXIMATELY HERE**

By utilizing the $ZIP_h$ model we observe that rainfall, average, maximum and minimum temperature effects are significant predictors of the Poisson component. Furthermore, autumn seasonal effects are detected for the Poisson component. Also the presence of autumn and summer seems to affect negatively the non-occurrence of the disease, as indicated by the negative sign of posterior parameter of the specific covariate in the $ZIP_h$ model including modeling of excess zero counts. The number of villages infected during the previous week is also important and negatively affects excess zero counts.

## 4. Discussion

As in *Choi et al. (2012)*, we used the previous week's covariate information to predict the present week's count of sheep pox cases because the present week's covariate information could not be available until the end of the week. *Choi et al. (2012)* suggest that the effects of temperature, humidity and rainfall are not significant and attribute this to two reasons: (i) the fact that the mechanism disease transmission may be more akin to an animal-to-animal contact spread than spreading via contaminated soil, for example, where environmental factors could be more important, and (ii) the monthly measurements being too crude to describe short term variations which might be better described by weekly data. The environmental factors included in the current analysis were shown to be significant in predicting cases of sheep pox. The weekly nature of our data may partly assist this outcome.

Inspecting each covariate separately reveals that the number of expected sheep pox cases in each week increases with the number of villages affected the previous week. Since that the number of villages affected in the previous week can be considered as a measure of the spatial dispersion of disease we have a strong indication that a greater dispersion of the contaminated load of the virus leads to an increase in the number of occurrences. Examining the signs of the weekly average, maximum and minimum temperatures shows that a decrease in the average temperature indicates an increase in the sheep pox counts while the maximum and minimum temperatures have the opposite effect. This is an interesting finding, indicating that the movement from low to high temperatures probably reduces incidence of sheep pox, up to a certain level of very high temperatures, beyond which the virus becomes again very effective. Average weekly precipitation levels, found significant only for the $ZIP_h$ model, seem to negatively affect sheep pox counts. The meteorological factor of relative humidity (coefficient $β_6$) was not significant in any of the fitted models, indicating perhaps the relative importance of direct/indirect contact spread in comparison to airborne spread. Generally, it seems that the weekly nature of the collected data, combined with explicitly modeling the disease counts via appropriate distributions, assisted in uncovering the importance of meteorological information.

The findings of our study, especially those concerning meteorological effects are not uncommon in the literature. For instance, high temperature effects were also found in *Bhanuprakash et al. (2005)* and *Webster (1981)*. Also, lower levels of rainfall were deemed significant, reducing disease occurrence (*Bhanuprakash et al., 2005*), as outlined in the current study by the $ZIP_h$ model. Rainfall appears to reduce disease spread for a number of reasons. This may be attributed to heavy rainfall preventing infectious virus particles from being carried in the air for long distances, thus infection can no long spread rapidly (*Bhanuprakash et al., 2005*). In addition, the presence of rainfall reduces grazing.

As *Bhanuprakash et al. (2006)* stress, sheep pox has been reported to occur throughout the year depending upon season, climate, specific agent of sheep pox virus and on various characteristics of sheep (see also *Achour and Bouguedour, 1999*). However, studies have associated high frequencies of sheep pox with unusual winter conditions (*Hailat et al., 1994*), and drought (*Chamoiseau, 1985*). Peak outbreaks have been reported during winter and summer months and less during August in certain parts of India. Therefore, it is evident that the relevant literature on the epidemiology of sheep pox indicates seasonality in the disease appearance (e.g. *Yeruham et al., 2007; Bhanuprakash et al., 2005*). To investigate the validity of temporal effects we have included in our model covariate information related to the season of the year. Seasonality effects were found to be significant for all fitted models, specifically autumn season is statistically important in the Poisson and negative binomial models and their zero-inflated extensions [ZIP and ZINB], whereas summer effects were also observed to affect non-occurrence of the disease in the ZIP$_h$ model.

The positive/negative coefficient for autumn covariate for the Poisson and the excess zero component in the ZIP$_h$ model, respectively, suggests that during this season the counts are expected to increase. This finding clearly indicates that in cases of future similar incidents, measures for coping with the disease should be more targeted during the autumn season. For example, authorities could pose more strict preventive measures such as intensified restriction measures concerning grazing of animals in the open field. *Mangana et al. (2008)* attribute the lack of spring cases to the high summer temperatures whence dry scabs separate and fall off the animals, a phenomenon which may assist in the spread of the disease through viral particles that can be easily carried via cars or people to distant regions. The negative statistically important coefficient $\beta_8^z$ for the ZIP$_h$ model suggests that during summer we have a reduction in the non-occurrence of the disease.

Regarding sheep pox seasonality, our study deviates from the literature where increasing disease incidents are primarily associated with winter rather than autumn months (*Yeruham et al., 2007; Hailat et al., 1994*). Another characteristic of the disease noted in the relevant literature is the one of seasonal variations of outbreaks of sheep pox. Specifically, *Hailat et al. (1994)* associate a sheep pox outbreak in Jordan with unusual winter conditions at a particular time period (January – February, 1992). In Israel *Yeruham et al. (2007)* report that two-thirds of the outbreaks have occurred during the winter and spring months (November to May), and explain this seasonality by the ability of the virus to persist for many months in wet and cold weather in association with the lambing season (*Garner et al., 2000*), or by the poor physiological condition of flocks in the autumn (*Achour et al., 2000*). *Bhanuprakash et al. (2005)* also find evidence of temporal behaviour of the disease in India, with highest number of outbreaks occurring in March and the fewest in August. The peak of disease between November and May might be due to the fact that during this period sheep are exposed to adverse temperatures, causing the suppression of their immune system therefore leading them to infection. Seasonality in the occurrence of sheep pox has also been reported in other studies, for instance in *Kitching et al. (1986), Le Jan et al. (1987),* and *Achour and Bougnedour (1999)*, where in all the aforementioned cases the occurrences were between a period between October and July. However, autumn effects have also been identified as significant in predicting disease counts by *Choi et al. (2012)*, who examined a similar epidemic phase for foot and mouth disease. Nonetheless, we must note here that comparisons of our study with the aforementioned literature should be approached with caution, since those analyses refer to different environmental conditions (e.g., typical winter temperatures in countries like India are more or less comparable with autumn temperatures in the region of Evros). Posterior estimates of correlation coefficients, ρ, for the OU process were very high for the ZIP and ZINB models and comparatively low for the Poisson model. Finally, number of occurrences in the previous week was found to be marginally important for the prediction of the occurrences in the current week. The latter is not unexpected though, since that our models include three components of autocorrelation, namely the covariate for the number of villages with infections in the previous week, the OU process taking into account the residual autocorrelation and the $\tau$ parameter which is taken out by the model. The conducted analysis indicated $ZIP_h$ as the best model to describe sheep pox epidemic, since that the specific structure of the collected data is a typical example of excess zeros count data.

To conclude, by utilizing sheep pox historical data using Bayesian methodology we have provided modeling strategies for the analysis of epidemiological data on animal diseases. We have shown that the sheep pox data strongly support the Poisson-based models as opposed to those stemming from the negative binomial distribution. From our analysis it appears that seasonality plays an important role in the frequency of sheep pox occurrence. Other useful predictors included the number of villages affected in the previous week, as well as specific environmental/meteorological variables, with the most important being the average, maximum and minimum temperatures and secondly rainfall.

We believe that our study sheds some light regarding future epidemiological studies and control efforts, especially in similar environmental conditions. Results and country-specific parameter estimates obtained from our epidemic models could be also used to predict the spread of disease in case of the virus is re-introduced into the country, by applying suitable simulation modeling schemes. In this way, one could provide additional aid to decision makers (government/livestock agencies) in their effort to control and evidently eradicate the disease.

Further, we have demonstrated that by fitting a simple model that relies on relatively little information on spread of the disease such as the crude measurement of spatial behavior based on number of villages affected in previous week, instead of complex models that require parameterizing in much detail the mechanism of disease spread, one may obtain an adequate estimation of the spatial and temporal behavior of the spread.

Finally, the consideration of mixture distributions and stochastic components into the modeling of the disease process resulted in improved model fit, suggesting that future modeling approaches for predicting similar epidemic episodes should consider such an approach. We are currently investigating the application of this class of models to alternative diseases such as foot and mouth.

**Tables**

**Table 1:** Number of sheep pox cases and total animal deaths during period 1994-98 in Evros Prefecture, Greece by month

| Month/Year | Number of cases | Number of Deaths |
|---|---|---|
| 11/94 | 2 | 83 |
| 10/95 | 6 | 416 |
| 11/95 | 2 | 261 |
| 7/96 | 6 | 666 |
| 8/96 | 3 | 395 |
| 9/96 | 28 | 2,772 |
| 10/96 | 34 | 6,651 |
| 11/96 | 49 | 8,373 |
| 12/96 | 15 | 3,853 |
| 1/97 | 1 | 110 |
| 8/97 | 11 | 1,727 |
| 9/97 | 20 | 3,165 |
| 10/97 | 10 | 2,085 |
| 11/97 | 7 | 913 |
| 12/97 | 5 | 948 |
| 7/98 | 12 | 1,618 |
| 8/98 | 2 | 240 |
| 9/98 | 14 | 546 |
| 10/98 | 14 | 388 |
| 11/98 | 6 | 232 |
| **TOTAL** | **249** | **35,442** |

**Table 2:** DIC, posterior mean deviance ($\overline{D}$) and posterior mean deviance minus deviance evaluated at the posterior mean of the parameters ($p_D$) for the 5 models

| Model | $\overline{D}$ | $p_D$ | DIC |
|---|---|---|---|
| Poisson | 310.2 | 56.52 | 366.76 |
| Negative binomial | 367.5 | 45.36 | 412.9 |
| ZIP | 277.3 | --- | --- |
| ZIP$_h$ | 270 | --- | --- |
| ZINB | 401.6 | --- | --- |

**Table 3:** Posterior medians and corresponding 95% credible intervals of the final best selected model (ZIP$_h$) parameters

| Parameter | Poisson component | Excess zero component |
|---|---|---|
| intercept ($\beta_0$) | ----- | **3.937** |
| | | **(1.679,7.334)** |
| # of infected villages in the previous week ($\beta_1$) | ----- | **-0.994** |
| | | **(-1.73,-0.357)** |
| rainfall ($\beta_2$) | **-0.061** | ----- |
| | **(-0.111,-0.017)** | |
| average temperature ($\beta_3$) | **-0.327** | ----- |
| | **(-0.593,-0.057)** | |
| maximum temperature ($\beta_4$) | **0.217** | ----- |
| | **(0.051,0.386)** | |
| minimum temperature ($\beta_5$) | **0.161** | ----- |
| | **(0.033,0.296)** | |
| summer ($\beta_8$) | ----- | **-2.072** |
| | | **(-3.638,-0.63)** |
| autumn ($\beta_9$) | **0.535** | **-3.152** |
| | **(0.174,0.908)** | **(-4.9,-1.531)** |
| *τau* | **0.076** | ----- |
| | **(0.059,0.105)** | |
| *sigma* | **0.018** | **1.718** |
| | **(0.001,0.097)** | **(0.246,6.196)** |
| *rho* | **0.999** | **0.853** |
| | **(0.998,0.9996)** | **(0.108,0.988)** |

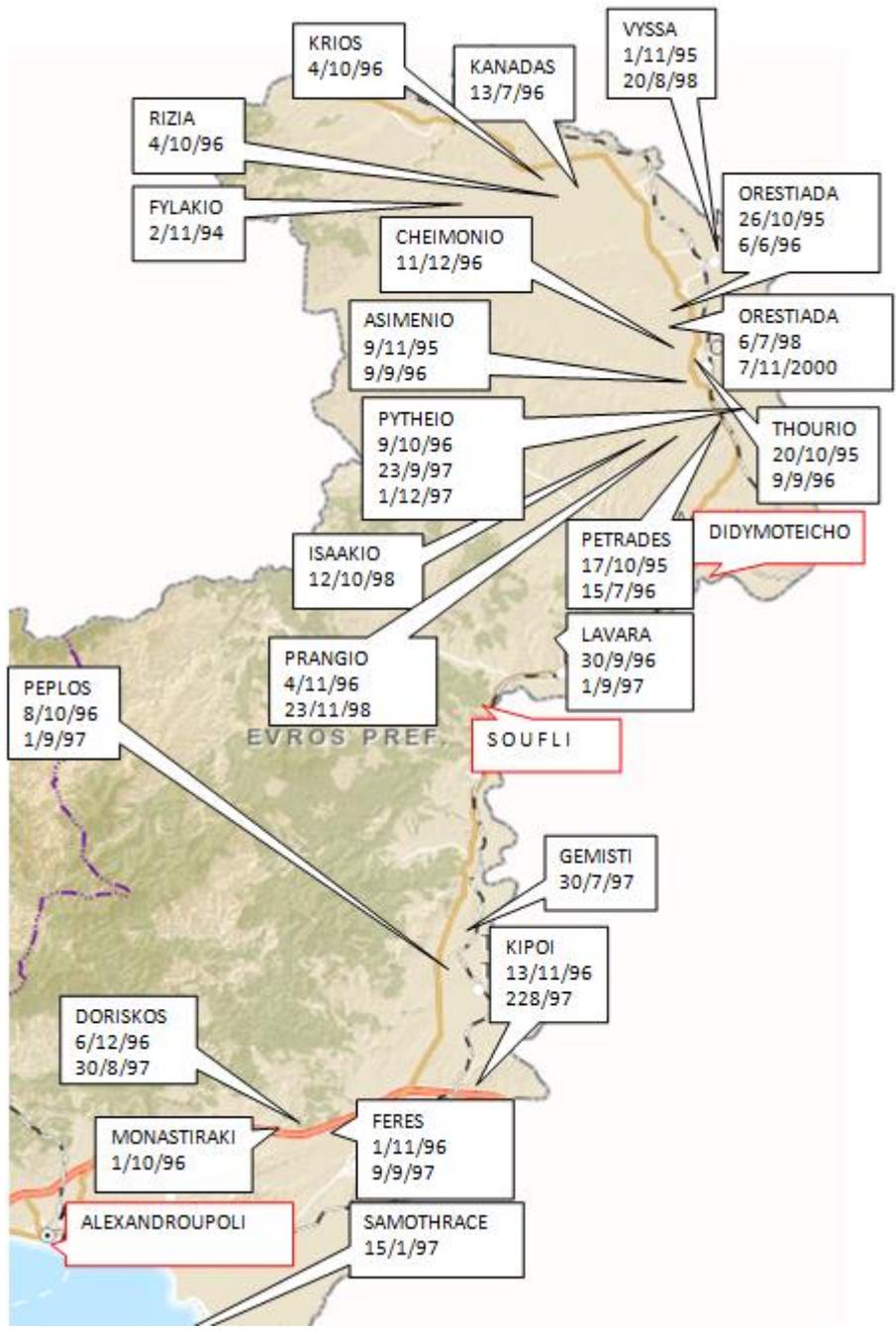

**Figure 1:** Map of the course of sheep pox epidemic in Evros Prefecture, Greece (1994-1998 & single case of 2000, ——— villages with infection; ——— 3 largest cities of the prefecture except Orestiada)

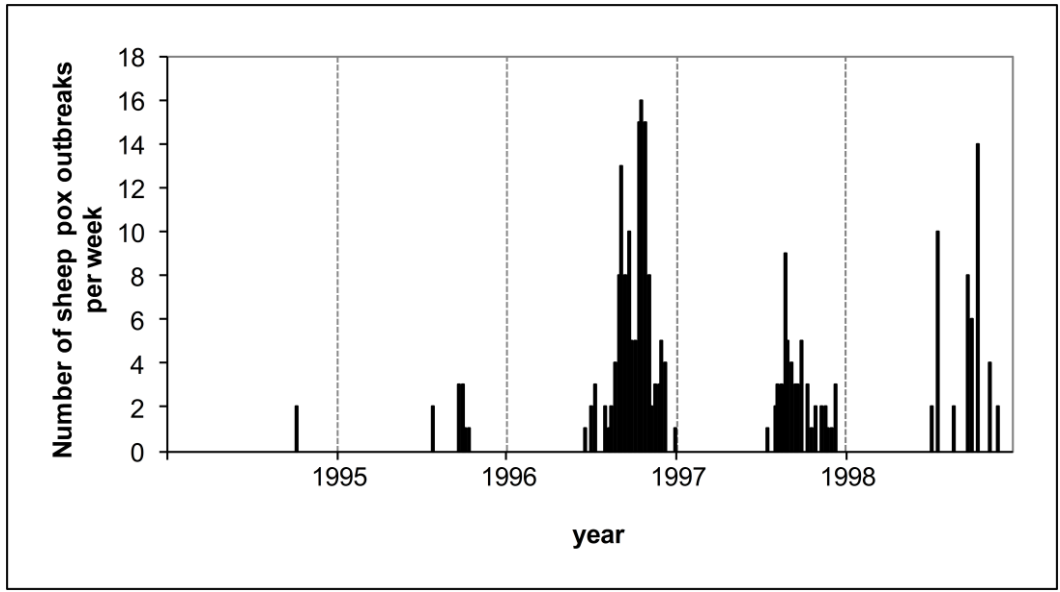

**Figure 2:** Temporal distribution of sheep pox cases in Evros, Greece (1994-1998)

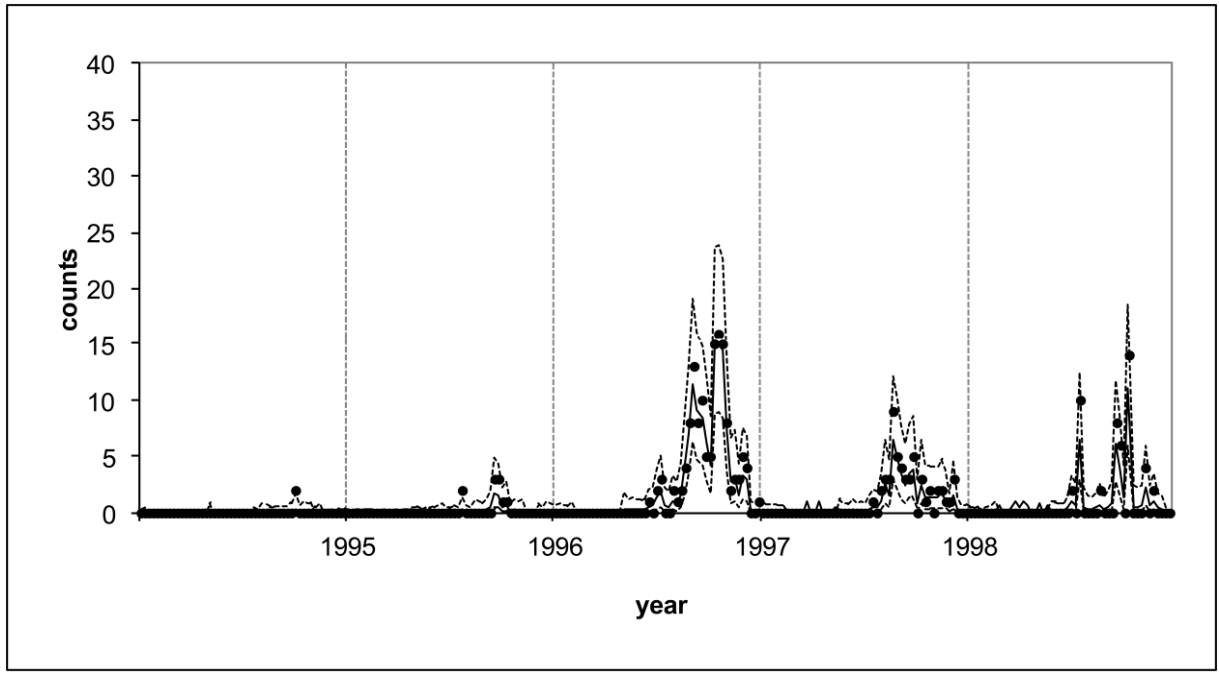

**Figure 3a:** Predicted vs observed counts for the Poisson model (•, observed; - - - - -, 95% probability intervals; ——— , predicted).

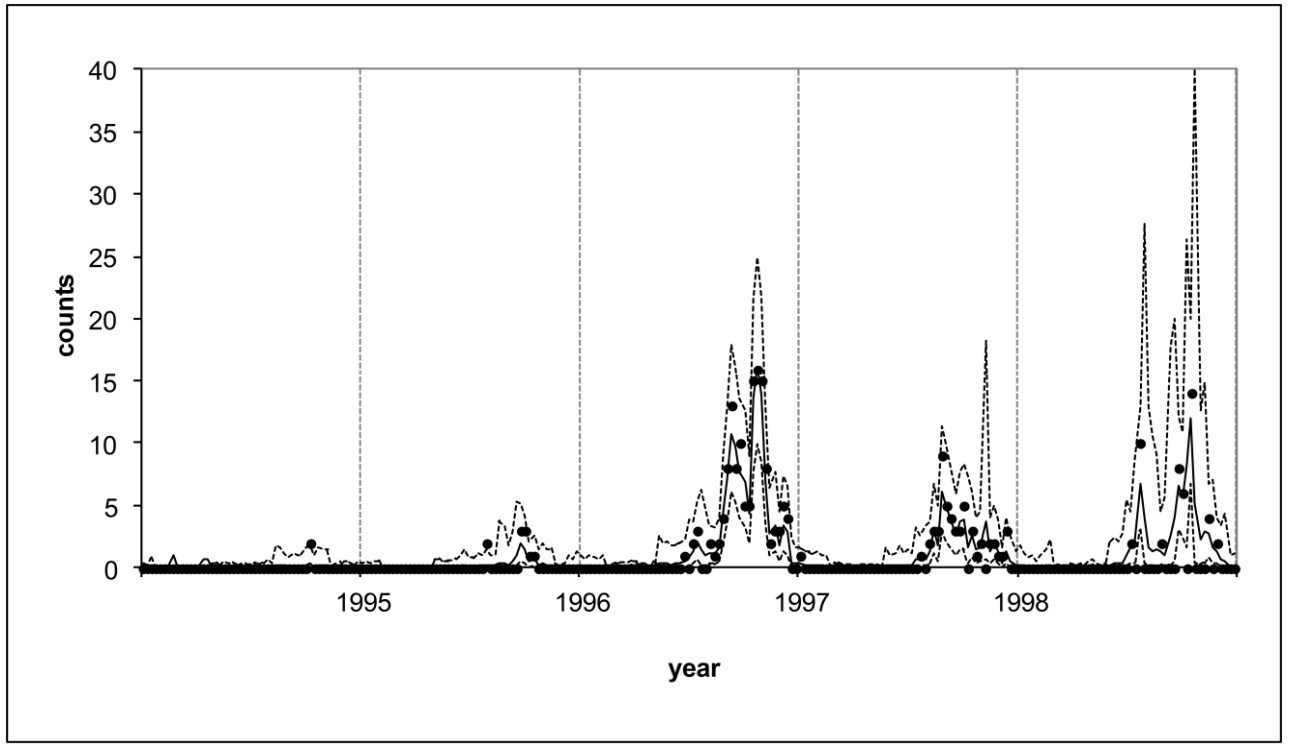

**Figure 3b:** Predicted vs observed counts for the ZIP model (•, observed; - - - - -, 95% probability intervals; ——— , predicted).

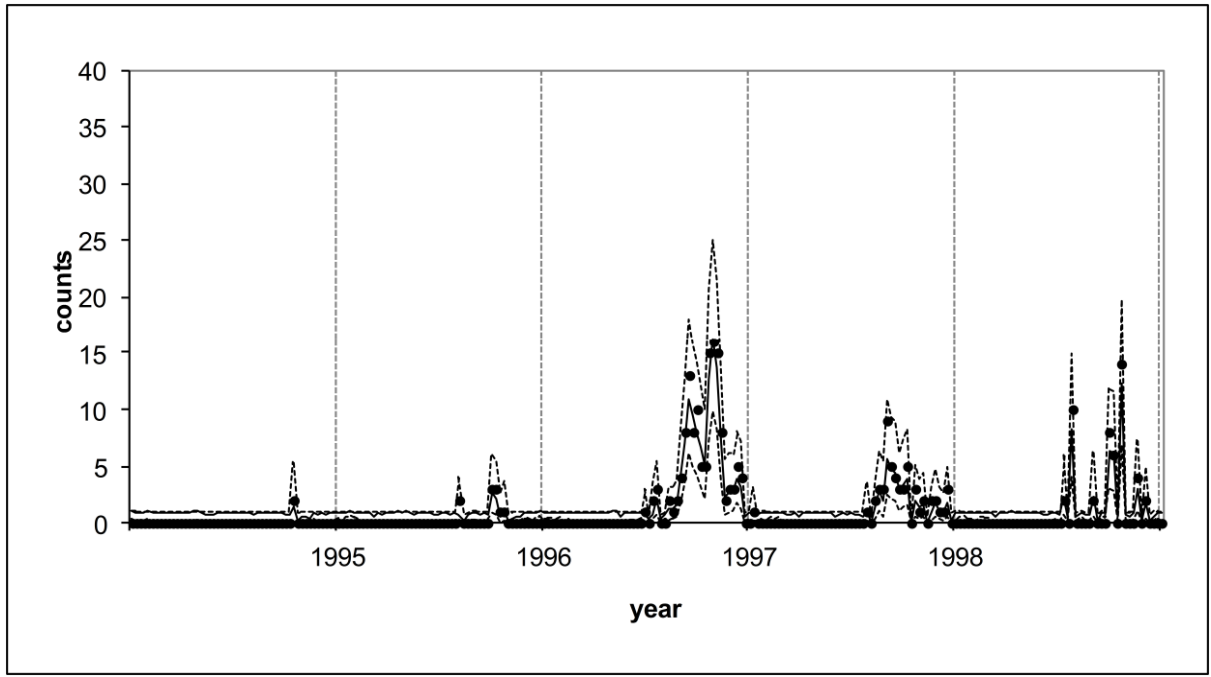

**Figure 3c:** Predicted vs observed counts for the $ZIP_h$ model (•, observed; - - - - -, 95% probability intervals; ———, predicted).